\documentclass[%
 reprint,
 amsmath,amssymb,
 aps,
]{revtex4-1}

\def\L{\bm{\mbox{-----}}}
\def\dashL{\bm{\mbox{--~--~--}}}
\def\dotL{\bm{\mbox{$\cdot\ \cdot\ \cdot$}}}

\usepackage{graphicx}
\usepackage{dcolumn}
\usepackage{bm}
\usepackage{subfigure}
\usepackage{xcolor}
\usepackage{soul}

\begin{document}

\title{The Effect of Discrete Resonant Manifold Structure on Discrete Wave Turbulence}

\author{Alexander Hrabski}
\author{Yulin Pan}%
 \email{yulinpan@umich.edu}
\affiliation{%
Department of Naval Architecture and Marine Engineering, University of Michigan, Ann Arbor, Michigan 48109, USA
}

\date{\today}

\begin{abstract}
We consider the long-term dynamics of nonlinear dispersive waves in a finite periodic domain. The purpose of the work is to show that the statistical properties of the wave field rely critically on the structure of the discrete resonant manifold (DRM). To demonstrate this, we simulate the two-dimensional MMT equation on rational and irrational tori, resulting in remarkably different power-law spectra and energy cascades at low nonlinearity levels. The difference is explained in terms of different structures of the DRM, which makes use of recent number theory results. 

\end{abstract}

\maketitle

\section{Introduction}
\vspace{-0.3cm}

Wave turbulence describes the statistical behavior of a large number of dispersive waves under nonlinear interactions. For any given physical context described by nonlinear wave equations, wave turbulence theory (WTT) predicts an inertial-range power-law spectrum associated with an energy cascade process, analogous to the Kolmogorov description of hydrodynamic turbulence. Due to the generality of this mathematical framework, WTT has found wide applications including surface and internal gravity waves \cite[e.g.][]{Zakharov1968,YLvov2010}, capillary waves \cite[e.g.][]{Zakharov1967}, plasma waves \cite[e.g.][]{Galtier2000}, acoustics \cite[e.g.][]{Lvov1997} and gravitational waves \cite[e.g.][]{Galtier2017}. 

In spite of its success, WTT relies on the assumption of an infinite domain, which is violated in finite experimental facilities and computational domains (say, with periodic boundary conditions). In these systems, the wave dynamics is altered by the discreteness in wave number, $\Delta k$, that results from the boundary conditions of the finite domain. Together with other mechanisms such as broad-scale dissipation \cite{Miquel2014,Deike2013,Deike2014,Pan2015} and bound wave generation \cite{Cobelli2011,Michel2018,Campagne2019}, the discreteness $\Delta k$ results in deviations of spectral slope and energy flux from WTT solutions \cite[e.g.][]{Pushkarev2000,YLvov2006,Pan2014,Denissenko2007,Hassaini2018,Deike2015,Nazarenko2010,Cazaubiel2019,Miquel2011,Mordant2010}. The effect of $\Delta k$ on wave dynamics can be understood through the quasi-resonance conditions (say, for a quartet)
\begin{equation}
\begin{split}
\bm{k}_{1}+\bm{k}_{2}-\bm{k}_{3}-\bm{k}  &=0,\\
\vert\omega_{\bm{k}_{1}}+\omega_{\bm{k}_{2}}-\omega_{\bm{k}_{3}}-\omega_{\bm{k}}\vert &\leq \Omega,
\label{eqn:res_cond}
\end{split}
\end{equation}
where $\bm{k}$ represents the wave number vector and $\omega$ is the angular frequency determined from $\bm{k}$ by the dispersion relation. $\Omega$ is the nonlinear broadening in frequency, which increases with nonlinearity level \cite[e.g.][]{Pan2017,Lvov2010}. It is postulated that the WTT spectrum can only be (approximately) obtained in the regime of kinetic wave turbulence (KWT), where the nonlinear broadening overcomes the discreteness, i.e., $\Omega > \Delta \omega$, with $\Delta \omega$ the frequency discreteness associated with $\Delta k$. However, the rigorous treatment of the KWT regime is still an open question despite many recent theoretical works \cite{Eyink2012,Chibbaro2017b,Chibbaro2018,Dymov2019,Deng2019}, especially regarding how to take the limits of infinite domain and weak nonlinearity by keeping $\Omega > \Delta \omega$. For $\Omega < \Delta \omega$, the regime of discrete wave turbulence (DWT), the number of interacting quartets of the wave field is reduced, resulting in a deviation from the dynamics predicted by WTT. Both experiments and numerical simulations of DWT have shown, in many cases, steepened spectra compared to the KWT regime \cite{Pushkarev2000,Pan2014,Denissenko2007,Hassaini2018,Deike2015,Denissenko2007}, as well as restricted energy transfer \cite{YLvov2006,Connaughton2001,Cazaubiel2019,Mordant2010} or ``frozen turbulence'' \cite{Pushkarev2000,Miquel2014} for sufficiently small $\Omega$. The transition from the DWT to the KWT regime is associated with an increasing nonlinearity level which triggers a ``spectral avalanche'' described in the ``sandpile'' model \cite{Nazarenko2006}.

The traditional sandpile description implies that the deviation of bounded domain dynamics from WTT depends only on the nonlinearity level and wave number discreteness. In the present work, we show that this traditional understanding overlooks a key property - the discrete resonant manifold (DRM) that survives as nonlinear broadening $\Omega$ approaches zero, i.e., the discrete set of resonant quartets which satisfy \eqref{eqn:res_cond} with $\Omega=0$. To illustrate this idea, we conduct controlled numerical simulations where the effect of wave number discreteness (thus the DRM) on dynamics can be studied separately, i.e., isolated from the other factors of broad-scale dissipation and bound waves. 

\begin{figure*}
  \centering
  \subfigure{\includegraphics[width=8.6cm]{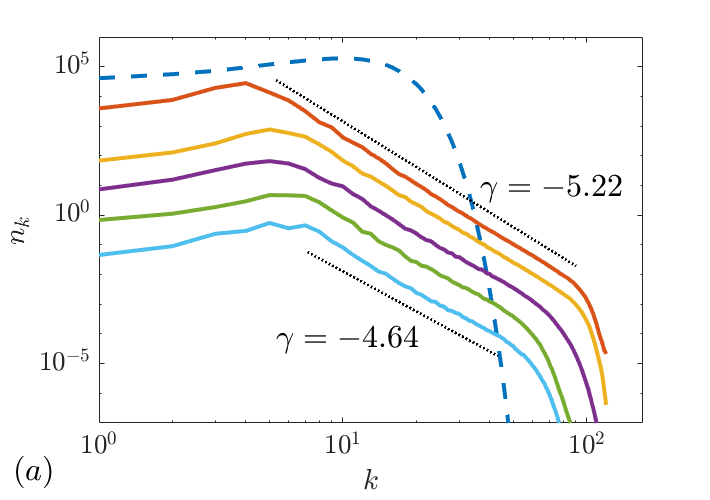}}
  \subfigure{\includegraphics[width=8.6cm]{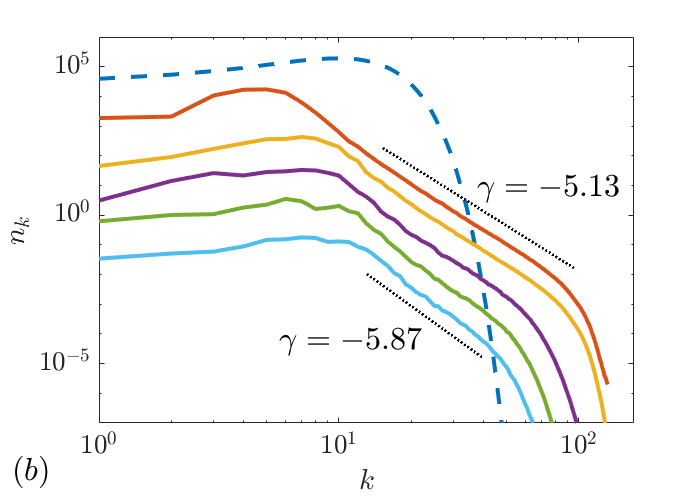}}
\caption{The initial spectra (\dashL) and fully-developed power-law spectra $n_k$ (\L) calculated with $\nu_{opt}$ at different values of $\epsilon$ on (a) $\mathbb{T}^2_r$ and (b) $\mathbb{T}^2_{ir}$. The power-law spectra are shifted for clarity, representing $\epsilon$=$0.03$, $0.01$, $0.003$, $0.001$ and $0.0003$ from top to bottom. The linear fit for the top and bottom spectra are indicated (\dotL).}
\label{fig:spectra}
\end{figure*}

The system under present study is a two-dimensional Majda, McLaughlin, Tabak (MMT) equation \cite{Majda1997,Cai1999,Zakharov2001}, which is chosen due to its generality in representing nonlinear dispersive wave dynamics and effectiveness in displaying many features relevant to wave turbulence \cite{Cai1999,Chibbaro2017}. The parameter in our MMT equation is tuned to yield a dispersion relation $\omega=|\bm{k}|^2$, only for which a rigorous interpretation of the simulation results can be achieved through available number theory results \cite{Faou2016}. The MMT equation is simulated on both rational and irrational tori (periodic domains of rational and irrational aspect ratios), which correspond to different DRM structures. We show remarkably different power-law spectra on these two tori, with the rational-torus spectral slope approaching the WTT solution with decreasing nonlinearity, in contrast to the steepened spectrum on the irrational torus. The dynamical differences between the tori are interpreted through the different DRM structures, over which a summation of resonant interactions critically determines the energy cascade. It is found that the DRM approximates the continuous resonant manifold (CRM) of WTT (i.e., the continuous set satisfying \eqref{eqn:res_cond} with $\Omega=0$) only on the rational torus. This can be tied to a recent number theoretic result \cite{Faou2016}, which rigorously equates the lattice sum over the DRM of the rational torus to the integration over the CRM with a constant factor difference. We conclude by outlining the implications of our findings to general physical wave contexts.

\vspace{-0.2cm}
\section{Numerical Setup}
\vspace{-0.2cm}

We consider the the two-parameter $(\alpha,\beta)$ MMT equation in two spatial dimensions:
\begin{equation}
i\frac{\partial \psi}{\partial t}=\vert\partial_{\bm{x}}\vert^{\alpha}\psi+
\vert\partial_{\bm{x}}\vert^{-\beta/4}\left(
\left|\vert\partial_{\bm{x}}\vert^{-\beta/4}\psi\right|^{2}\vert\partial_{\bm{x}}\vert^{-\beta/4}\psi \right),
\label{eqn:MMT}
\end{equation}
where $\psi\equiv\psi(\bm{x},t)$ with $\bm{x}$ the spatial coordinates and $t$ the time. $\vert\partial_{\bm{x}}\vert^{\alpha}$ denotes a multiplication by $k^{\alpha}$ on each spectral component in wave number domain, with $k=|\bm{k}|$. The parameter $\alpha=2$ is chosen, resulting in $\omega=k^2$, the dispersion relation as the nonlinear Schrödinger equation. We choose $\beta=-4$, which corresponds to fast spectral evolution with nonlinear time scale decreasing with the increase of $k$ \cite{Falkovich1991}. The MMT equation \eqref{eqn:MMT} corresponds to a Hamiltonian system with the Hamiltonian $H=H_0+H_1$, with linear and nonlinear components $H_0$ and $H_1$ provided in \cite{Supp}.

Application of WTT on \eqref{eqn:MMT} (in an infinite domain) yields an analytical solution of isotropic spectrum 
\begin{equation}
    n_k = n_{\bm{k}} \equiv \langle \hat{\psi}_{{\bm{k}}} \overline{\hat{\psi}_{{\bm{k}}}} \rangle \sim k^{\gamma_0},
    \label{eqn:MMT_theory}
\end{equation}
where $\gamma_0=-14/3$, $\hat{\psi}_{{\bm{k}}}$ is the Fourier transform of $\psi$, $\overline{\hat{\psi}_{{\bm{k}}}}$ denotes the complex conjugate of $\hat{\psi}_{{\bm{k}}}$, angle brackets denote an ensemble average, and $n_k$ is the angle-averaged wave action spectrum. This isotropic spectrum represents a stationary solution of the WTT kinetic equation \cite{Zakharov2001,Supp}
\begin{equation}
\begin{split}
\frac{\partial n_{\bm{k}}}{\partial t}=&\underset{(\bm{k}_{1},\bm{k}_{2},\bm{k}_{3})\in \mathbb{R}^2}{\int4\pi k^{2} k_{1}^{2}} \!\! k_{2}^{2} k_{3}^{2} (n_{\bm{k}_{1}}n_{\bm{k}_{2}}n_{\bm{k}_{3}}+n_{\bm{k}_{1}}n_{\bm{k}_{2}}n_{\bm{k}} \\
-n_{\bm{k}_{1}}n_{\bm{k}_{3}}&n_{\bm{k}}-n_{\bm{k}_{2}}n_{\bm{k}_{3}}n_{\bm{k}}) \delta\left(\bm{k}_{1}+\bm{k}_{2}-\bm{k}_{3}-\bm{k}\right) \\
&\delta\left(\omega_{\bm{k}_{1}}+\omega_{\bm{k}_{2}}-\omega_{\bm{k}_{3}}-\omega_{\bm{k}}\right)d\bm{k}_{1} d\bm{k}_{2} d\bm{k}_{3},
\label{eqn:KE}
\end{split}
\end{equation}
where $\delta$ is the Dirac delta function. 

In order to study the effect of DRM structure, we also consider \eqref{eqn:MMT} on rational and irrational tori $\mathbb{T}^2_r$ and $\mathbb{T}^2_{ir}$ of sizes $2\pi \times 2\pi / q$, with $q=1$ for $\mathbb{T}^2_r$ and $q^2=\sqrt{2}$ for $\mathbb{T}^2_{ir}$. The corresponding discrete wave numbers are taken from the sets $\mathbb{Z}^2_{r,ir} \equiv \mathbb{Z}\times q\mathbb{Z}$. While any irrational number $q$ results in an irrational torus, the particular value $q^2=\sqrt{2}$ eliminates the majority of resonant quartets that exist on $\mathbb{T}^2_r$ by restricting the orientation of quartets on the $k$-plane (see \cite{Supp} for a detailed discussion).

We simulate \eqref{eqn:MMT} using a GPU-accelerated pseudo-spectral method with $256\times256$ modes on both $\mathbb{T}^2_r$ and $\mathbb{T}^2_{ir}$. Multiple simulations of free-decay turbulence are conducted, starting from isotropic Gaussian spectra $n_{\bm{k}}=a_0e^{-0.01(k-10)^2}$ with random phases and amplitude $a_0$ covering a broad range of nonlinearity levels. Since the integral of energy described by \eqref{eqn:MMT_theory} is convergent in the limit of $k\rightarrow \infty$, the (theoretical) forward cascade forms a finite-capacity spectrum \cite{Nazarenko_WTTBOOK} which is realizable in free-decay simulations. To model wave dissipation, we add a hyper-viscosity term $-i\nu \vert\partial_{\bm{x}}\vert^{8}\psi$ to the right hand side of \eqref{eqn:MMT}. We remark that the $k^8$ dependence is sufficient to confine the dissipation at high wave numbers \cite{Majda1997,Chibbaro2017}, in contrast to some previous simulations \cite{Pan2015} and experiments \cite{Miquel2014,Deike2014}. The parameter $\nu$ takes values $\nu_R \equiv[2.50\times 10^{-17}, 2.50\times 10^{-16}]$, for which clear power-law spectra can be observed at all nonlinearity levels of interest. We further consider an optimal value $\nu_{opt}$ (for each nonlinearity level on each torus) that corresponds to the smallest $\nu$ in $\nu_R$ without resulting in ``bottleneck'' energy accumulation at high wave numbers. In the next section, we report results for $\nu_{opt}$ as well as uncertainty bars associated with $\nu_R$. 

\section{Results}
\vspace{-0.4cm}

We define the nonlinearity level of a wave field as $\epsilon\equiv H_1/(H_0+H_1)$. In figure \ref{fig:spectra}, the initial and fully-developed spectra $n_k$ at different values of $\epsilon$ obtained with $\nu_{opt}$ are plotted for $\mathbb{T}^2_r$ and $\mathbb{T}^2_{ir}$. We see that the fully-developed spectra exhibit power-law forms with inertial ranges of about 2/3$\sim$4/3 decades depending on $\epsilon$ on each torus. We evaluate the spectral slope $\gamma$ based on a linear fit of the inertial power-law range. 

\begin{figure}
\includegraphics[width=8.6cm]{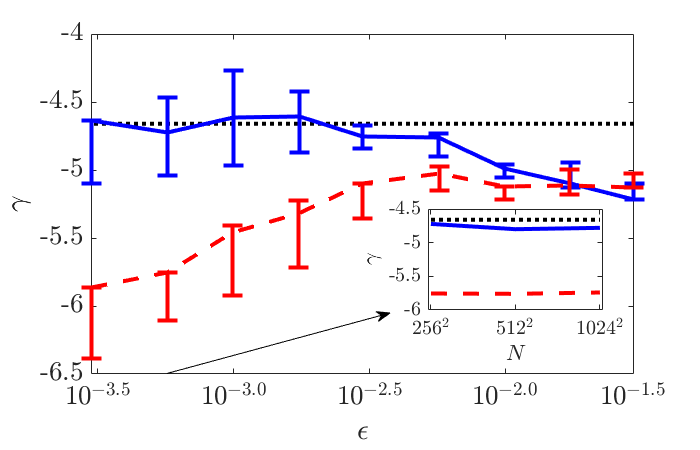}
\caption{Spectral slope $\gamma$ computed using $\nu_{opt}$ as a function of $\epsilon$ on $\mathbb{T}_r^2$ (\L) and $\mathbb{T}_{ir}^2$ (\dashL). The uncertainties associated with $\nu_R$ are shown by the vertical bars. The WTT analytical solution $\gamma_0=-14/3$ is indicated (\dotL). (inset) Spectral slope $\gamma$ computed with $\nu_{opt}$ at $\epsilon=5.75\times10^{-4}$ as a function of number of modes $N$ on $\mathbb{T}_r^2$ (\L) and $\mathbb{T}_{ir}^2$ (\dashL).}
\label{fig:slope_comp}
\end{figure}

\begin{figure}
  \centering
  \subfigure{\includegraphics[width=4.2cm,trim={0 2.5mm 0 0},clip]{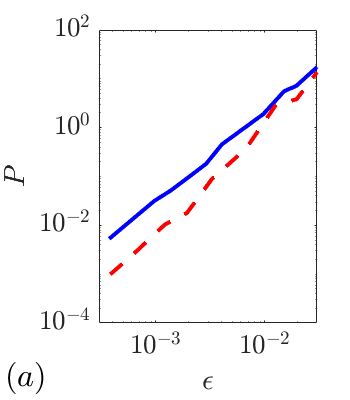}}
  \subfigure{\includegraphics[width=4.2cm,trim={0 2.5mm 0 0},clip]{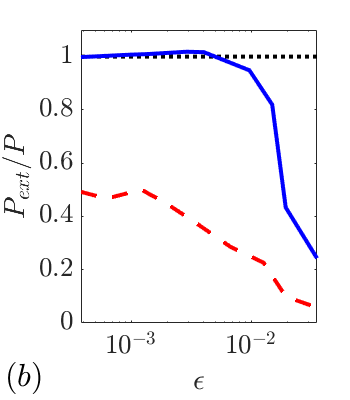}}
\caption{ (a) $P$ and (b) $P_{ext}/P$ as functions of $\epsilon$ on $\mathbb{T}_r^2$ (\L) and $\mathbb{T}_{ir}^2$ (\dashL). The line of $P_{ext}/P=1$ (\dotL) is indicated in (b).}
\label{fig:fluxes}
\end{figure}

\begin{figure*}
  \centering
  \subfigure{\includegraphics[width=5.6cm]{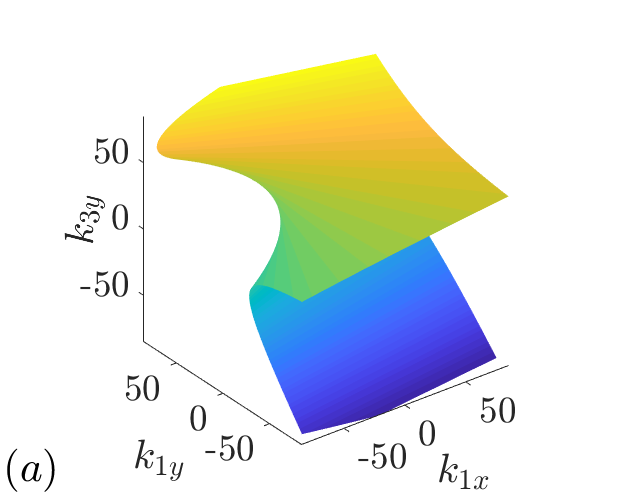}}
  \subfigure{\includegraphics[width=5.6cm]{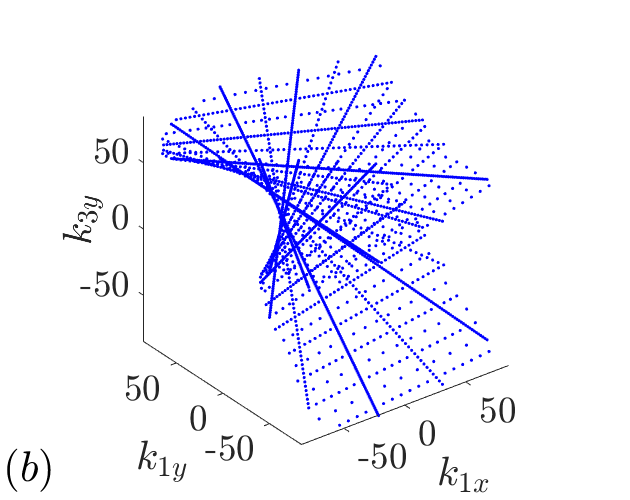}}
  \subfigure{\includegraphics[width=5.6cm]{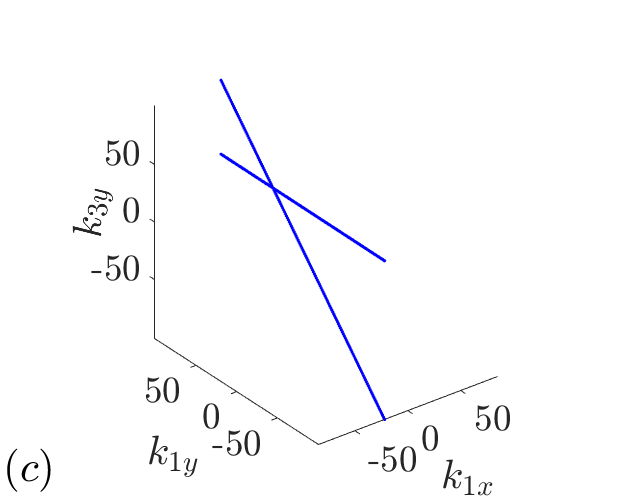}}
\caption{Visualizations of (a) the CRM, (b) the DRM on $\mathbb{T}_r^2$ and (c) the DRM on $\mathbb{T}_{ir}^2$ for $\bm{k}_2=(-36,31q)$ and $k_{3x}=-22$.}
\label{fig:res_set}
\end{figure*}

Figure \ref{fig:slope_comp} plots the spectral slope $\gamma$ obtained with $\nu_{opt}$ as a function of $\epsilon$ (ranging two orders of magnitude) on both tori. We also include the uncertainty bars computed from the range $\nu_R$, which show insignificant impacts on the values of $\gamma$ (with an order of magnitude variation of $\nu$, the largest uncertainty in $\gamma$ is $\mathcal{O}(0.5)$). At high nonlinearity levels, the values of $\gamma$ on both tori are almost identical, and about 0.5 smaller (i.e., steeper spectra) than the WTT value $\gamma_0=-14/3$. With the decrease of nonlinearity, the spectral slope $\gamma$ exhibits remarkably different behaviors on the two tori. On $\mathbb{T}^2_{ir}$, $\gamma$ decreases with $\epsilon$, indicating steepened spectra in agreement with previous observations in other wave systems \cite{Pushkarev2000,Pan2014,Denissenko2007,Hassaini2018,Deike2015,Denissenko2007}. However, on $\mathbb{T}^2_r$, $\gamma$ approaches and remains at $\gamma_0$ with the decrease of $\epsilon$, a trend unexplained by the traditional sandpile model. Convergence to the WTT spectral slope with the decrease of nonlinearity level, on the other hand, is also recently observed for a one-dimensional MMT equation \cite{Chibbaro2017} (whose resonant set may be considered as a subset of the 2D case without the depleting effect of $q$). To ensure the robustness of these results to varying grid resolutions, we select two cases at relatively low nonlinearity level on $\mathbb{T}_r^2$ and $\mathbb{T}_{ir}^2$, and check the spectral slope with increasing number of modes $N$ in simulations. The results plotted in the inset of figure \ref{fig:slope_comp} show that $\gamma$ converges with the increase of $N$ and that the results using $256^2$ modes are sufficient to capture the physics of interest.

The behavior of $\gamma$ cannot be interpreted by the bound wave mechanism as in some experiments \cite{Cobelli2011,Michel2018,Campagne2019}, as analysed in detail from the $k$-$\omega$ spectrum in \cite{Supp}. To understand the dominating mechanism, we further investigate the energy cascade on both tori. Unlike the previous evaluation of energy flux $P$ based on the dissipation (or energy input) rate \cite[e.g.][]{Pan2014,Pan2015,Falcon2007,Deike2014,Miquel2014}, we develop a new approach to evaluate $P$ directly from the dynamical equation \eqref{eqn:MMT} via a sum over interacting quartets, which rules out the uncertainties associated with the quasi-stationary state and the artificial form of the dissipation. The new approach also allows us to distinguish the contribution to $P$ from exact-resonant and quasi-resonant interactions, i.e., $P=P_{ext}+P_{qua}$, a key element of our analysis. From a control volume argument, we obtain
\begin{equation}
    P_*=-\sum_{k<k_b}{\omega_{\bm{k}}\frac{\partial n_{\bm{k}}}{\partial t}} \bigg|_*,
\end{equation}
where $*$ denotes either $ext$ or $qua$ (i.e., quantities associated with exact resonance or quasi-resonance), and $k_b$ is a wave number in the inertial range. We use $k_b=25$ for the subsequent results without loss of generality. The quantity $\partial n_{\bm{k}}/\partial t|_*$ can be directly derived from \eqref{eqn:MMT} \cite{Supp}:
\begin{equation}
\begin{split}
    \frac{\partial n_{\bm{k}}}{\partial t}\bigg|_* =
\sum_{(\bm{k}_{1},\bm{k}_{2},\bm{k}_{3})\in S_*} {2 k k_{1} k_{2} k_{3}}\text{Im}\langle\overline{\hat{\psi}_{\bm{k}}}\hat{\psi}_{\bm{k}_1}\hat{\psi}_{\bm{k}_2}\overline{\hat{\psi}_{\bm{k}_3}}\rangle, \\
\text{for}\ \bm{k}, \bm{k}_{1},\bm{k}_{2},\bm{k}_{3} \in \mathbb{Z}^2_r \ \text{or}\ \mathbb{Z}^2_{ir},
\end{split}
\label{eqn:dndt}
\end{equation}
where Im denotes the imaginary part of a function, $S_{ext}\equiv \{(\bm{k}_{1},\bm{k}_{2},\bm{k}_{3})|\bm{k}_{1}+\bm{k}_{2}-\bm{k}_{3}-\bm{k}=0, \  \omega_{\bm{k}_{1}}+\omega_{\bm{k}_{2}}-\omega_{\bm{k}_{3}}-\omega_{\bm{k}}=0\}$, and $S_{qua}\equiv \{(\bm{k}_{1},\bm{k}_{2},\bm{k}_{3})|\bm{k}_{1}+\bm{k}_{2}-\bm{k}_{3}-\bm{k}=0, \  \omega_{\bm{k}_{1}}+\omega_{\bm{k}_{2}}-\omega_{\bm{k}_{3}}-\omega_{\bm{k}}\neq 0\}$.

Figure \ref{fig:fluxes}(a) shows the total flux $P$ as a function of $\epsilon$ for both tori. The values of $P$ on $\mathbb{T}^2_{r}$ are consistently larger than those on $\mathbb{T}^2_{ir}$, with a pronounced difference (about one decade) for small $\epsilon$. This indicates that the energy cascade is far more efficient on $\mathbb{T}^2_{r}$, especially at low nonlinearity. We further plot $P_{ext}/P$ as a function of $\epsilon$ in figure \ref{fig:fluxes}(b).  With the decrease of $\epsilon$, it is found that $P_{ext}/P$ on $\mathbb{T}^2_{r}$ quickly approaches unity, showing the dominance of exact resonance on the energy cascade at low nonlinearity. The regime of $P_{ext}/P\approx 1$ occurs consistently with $\gamma\approx \gamma_0$ for nonlinearity level $\epsilon \lesssim 0.005$ on $\mathbb{T}_r^2$. On the other hand, the ratio $P_{ext}/P$ on $\mathbb{T}^2_{ir}$ increases much slower, not exceeding 50\% in the range of nonlinearities of interest. This analysis implies that the surviving exact resonances at low nonlinearity are critical in understanding the agreement between $\gamma$ and $\gamma_0$ on $\mathbb{T}^2_{r}$, as well as the steepened spectrum on $\mathbb{T}^2_{ir}$.

\section{Role of the DRM}
\vspace{-0.4cm}

In this section, we further investigate the structure of the resonant set $S_{ext}$, which becomes increasingly important to the dynamics with the decrease of nonlinearity level. In particular, we will show that $\partial n_{\bm{k}}/\partial t|_{ext}$ in \eqref{eqn:dndt} is related to the WTT kinetic equation, which explains $\gamma=\gamma_0$ at low nonlinearity on $\mathbb{T}^2_{r}$. 

To facilitate the description, we define the set $S_{ext}\cap \{ \bm{k}, \bm{k}_{1},\bm{k}_{2},\bm{k}_{3} \in \mathbb{T}_{r,ir}^2\}$ as the discrete resonant manifold (DRM) on $\mathbb{T}_{r}^2$ and $\mathbb{T}_{ir}^2$; and the set $S_{ext}\cap \{ \bm{k}, \bm{k}_{1},\bm{k}_{2},\bm{k}_{3} \in \mathbb{R}^2\}$ as the continuous resonant manifold (CRM) as in WTT. It has been rigorously proven by number theory \cite{Faou2016} that for the dispersion relation $\omega=k^2$, the summation in \eqref{eqn:dndt} over $S_{ext}$ (the DRM) on $\mathbb{T}^2_{r}$ converges to an integral on the corresponding CRM with a factor difference in the limit of high wave numbers. Built on the \emph{continuous resonant} equation given in \cite{Faou2016}, we can derive (see \cite{Supp} for details)
\begin{equation}
\begin{split}
\frac{\partial n_{\bm{k}}}{\partial t}\bigg|_{ext} \sim 
\underset{(\bm{k}_{1},\bm{k}_{2},\bm{k}_{3})\in S_{ext}}{\int2 k k_{1} k_{2} k_{3}} \! \!
\text{Im}\langle\overline{\hat{\psi}_{\bm{k}}}\hat{\psi}_{\bm{k}_1}\hat{\psi}_{\bm{k}_2}\overline{\hat{\psi}_{\bm{k}_3}}\rangle d\bm{k}_{1}d\bm{k}_{2}d\bm{k}_{3}, \\
\text{for}\ \bm{k}, \bm{k}_{1},\bm{k}_{2},\bm{k}_{3} \in \mathbb{R}^2.
\end{split}
\label{eqn:CR}
\end{equation}

At low nonlinearity, $\partial n_{\bm{k}}/\partial t=\partial n_{\bm{k}}/\partial t|_{ext}$ on $\mathbb{T}^2_{r}$. We can then formulate the kinetic equation \eqref{eqn:KE} (up to a factor difference) based on \eqref{eqn:CR}, under quasi-Gaussian statistics which are valid at low nonlinearity. Therefore, the spectral slope on $\mathbb{T}_r^2$ yields $\gamma=\gamma_0$ as in the stationary solution of \eqref{eqn:KE}. On the other hand, \eqref{eqn:CR} is not satisfied for $\mathbb{T}_{ir}^2$, resulting in the steepened spectra at low nonlinearity level as previously explained in the sandpile framework.

We further elaborate the structure of the DRMs on $\mathbb{T}_r^2$ and $\mathbb{T}_{ir}^2$ using a numerical visualization. To avoid viewing high-dimensional manifolds, we fix $\bm{k}_2=(-36,31q)$ and $k_{3x}=-22$, such that $S_{ext}$ is reduced to a two-dimensional structure embedded in a higher-dimensional space. Without loss of generality, we consider the higher-dimensional space to be $\{k_{1x},k_{1y},k_{3y}\}$, and plot the reduced CRM, and DRM on $\mathbb{T}_r^2$ and $\mathbb{T}_{ir}^2$ in figure \ref{fig:res_set}. While the DRM on $\mathbb{T}_r^2$ resembles the CRM, the DRM on $\mathbb{T}_{ir}^2$ is fundamentally different with a diminished number of resonant quartets. The salient contrast in these DRM structures is the inherent reason for different DWT dynamics on $\mathbb{T}_r^2$ and $\mathbb{T}_{ir}^2$. Finally, we remark that both the density \emph{and distribution} of resonances on the DRM affect the DWT dynamics. This is discussed in the contexts of other values of $q$ in \cite{Supp}, alongside a brief study on the effect of quasi-resonances under nonlinear broadening $\Omega$.

\vspace{-0.3cm}
\section{Conclusions and Discussions}
\vspace{-0.4cm}

Through simulations of the 2D MMT equation (with dispersion relation $\omega=k^2$) on rational and irrational tori ($\mathbb{T}_r^2$ and $\mathbb{T}_{ir}^2$), we identify the critical effect of the structure of discrete resonant manifold (DRM) on discrete wave turbulence (DWT). On $\mathbb{T}_r^2$, the DRM structure resembles the continuous resonant manifold (CRM), with a lattice summation over the DRM converging to an integral over the CRM. The spectral slope thus approaches $\gamma_0$ for low nonlinearity as predicted by the WTT kinetic equation. On $\mathbb{T}_{ir}^2$, the DRM is altered by the diminished number of resonant quartets, leading to steepened spectrum and reduced energy cascade capacity with the decrease of nonlinearity.

In general, the DRM structure relies on the physical wave properties and domain aspect ratio, with the former including dispersion relation and number of modes involved in each interaction. Therefore, different physical wave systems in different domains yield much richer DWT dynamics than that predicted by the sandpile model, which does not have sensitivity to these factors. For example, capillary wave turbulence exhibits steepened spectra with the decrease of nonlinearity level on $\mathbb{T}_r^2$ \cite{Pan2014}, exactly opposite to MMT turbulence. This is due to the different DRM of capillary waves, which yields an empty set on $\mathbb{T}_r^2$ \cite{Kartashova1990}. On the other hand, it is possible to use some irrational aspect ratio $q$ to restore some exact resonances for capillary waves, which may provide a system where the spectral slope is less sensitive to the nonlinearity level. Experimentally, the effect of $q$ can be observed by varying the aspect ratio of a wave tank as conducted in \cite{Hassaini2018}, although the effect of wall (instead of periodic) boundary conditions need to be addressed. A more effective experimental setting is wave turbulence on a fluid torus \cite{Laroche2019} which provides naturally periodic boundary conditions. Further, given the wide applications of WTT, more investigations are warranted for understanding the DWT dynamics from the perspective of DRMs for different physical wave systems, including the Fermi-Pasta-Ulam-Tsingou problem which is recently interpreted in the wave turbulence framework \cite{YLvov2018,Bustamante2019,Pistone2019}. 

This work also establishes connections between pure mathematics and the physics of wave turbulence. The number theoretic properties of the DRM, in terms of its relation to CRM by \eqref{eqn:dndt} and \eqref{eqn:CR}, allows the development of the first quantitative understanding of energy flux in DWT (through a new DWT kinetic equation). To extend the DWT kinetic equation to other physical wave systems, number theoretic problems concerning DRMs associated with other dispersion relations must be resolved. The present work is also of interest to mathematicians in the field of harmonic analysis, in particular for the quantification of energy cascades on tori. Our finding of diminished energy flux for the MMT equation at low $\epsilon$ on $\mathbb{T}_{ir}^2$ is indeed consistent with recent rigorous analysis of the nonlinear Schrödinger equation \cite{Staffilani2020}.

\vspace{-0.4cm}
\section{Acknowledgements}
\vspace{-0.3cm}

We would like to thank Professor Zaher Hani for fruitful discussions and his explanation of the number theoretic results. This research was supported in part through computational resources and services provided by Advanced Research Computing at the University of Michigan, Ann Arbor. This material is based upon work supported by the National Science Foundation Graduate Research Fellowship under Grant No. DGE 1841052.

\bibliography{citations}

\end{document}


\title{Supplemental Material for the Paper \\ ``The Effect of Discrete Resonant Manifold Structure on Discrete Wave turbulence''}

\author{Alexander Hrabski}
\author{Yulin Pan}
\affiliation{
Department of Naval Architecture and Marine Engineering, University of Michigan, Ann Arbor, Michigan 48109, USA
}

\maketitle

In this document, we provide:
\begin{enumerate}[I.]
    \item formulations regarding the two-dimensional (2D) MMT equation, the kinetic equation, and its analytical power-law solution in an infinite domain
    \item formulations regarding energy transfer associated with the MMT equation on a torus 
    \item formulation of the \emph{continuous resonant} system
    \item the configurations of the resonant quartets on $\mathbb{T}_r^2$ and $\mathbb{T}_{ir}^2$, including a range of $q$ and their effects on the spectral slopes
    \item quantification of nonlinear frequency broadening and visualization of the discrete quasi-resonant manifold (quasi-DRM)
\end{enumerate}

\section{MMT formulations in an infinite domain}

We start from the 2D MMT equation (parameters $\alpha=2$ and $\beta=-4$):
\begin{equation}
i\frac{\partial \psi}{\partial t}=\vert\partial_{\bm{x}}\vert^{\alpha}\psi+
\vert\partial_{\bm{x}}\vert^{-\beta/4}\left(
\left|\vert\partial_{\bm{x}}\vert^{-\beta/4}\psi\right|^{2}\vert\partial_{\bm{x}}\vert^{-\beta/4}\psi \right),
\label{MMT}
\end{equation}
which corresponds to a Hamiltonian system with Hamiltonian $H=H_{0}+H_{1}$, where 

\begin{equation}
\begin{split}
H_0&=\underset{\bm{x}\in \mathbb{R}^2}{\int}\left|\vert\partial_{\bm{x}}\vert^{\alpha/2}\psi\right|^{2}d\bm{x}, \\
H_1&=\underset{\bm{x}\in \mathbb{R}^2}{\int}\frac{1}{2}\left|\vert\partial_{\bm{x}}\vert^{-\beta/4}\psi\right|^{4}d\bm{x}.
\label{eqn:MMT_Ham}
\end{split}
\end{equation}

Taking the (non-unitary) Fourier transform $\psi\rightarrow \hat{\psi}_{\bm{k}}$ of \eqref{MMT}, we obtain the $k$-domain representation:
\begin{equation}
i\frac{\partial \hat{\psi}_{\bm{k}}}{\partial t}= k^{2}\hat{\psi}_{\bm{k}} +
\underset{\bm{k}_1,\bm{k}_2,\bm{k}_3 \in \mathbb{R}^2}{\int}k k_{1} k_{2} k_{3} \hat{\psi}_{\bm{k}_1}\hat{\psi}_{\bm{k}_2}\overline{\hat{\psi}_{\bm{k}_3}}
\delta\left(\bm{k}_{1}+\bm{k}_{2}-\bm{k}_{3}-\bm{k}\right)d\bm{k}_{1}d\bm{k}_{2}d\bm{k}_{3}.
\label{eqn:MMT_kdomain}
\end{equation}

Defining $\langle \hat{\psi}_{\bm{k}}\overline{\hat{\psi}_{\bm{k}'}}  \rangle = n_{\bm{k}} \delta (\bm{k}-\bm{k}')$, the evolution of $n_{\bm{k}}$ can be derived (through the subtraction of \eqref{eqn:MMT_kdomain}$\times\overline{\hat{\psi}_{\bm{k}}}$ from its conjugate): 
\begin{equation}
\frac{\partial n_{\bm{k}}}{\partial t}=
\underset{\bm{k}_1,\bm{k}_2,\bm{k}_3 \in \mathbb{R}^2}{\int}2 k k_{1} k_{2} k_{3}\text{Im}\langle\overline{\hat{\psi}_{\bm{k}}}\hat{\psi}_{\bm{k}_1}\hat{\psi}_{\bm{k}_2}\overline{\hat{\psi}_{\bm{k}_3}}\rangle 
\delta\left(\bm{k}_{1}+\bm{k}_{2}-\bm{k}_{3}-\bm{k}\right)d\bm{k}_{1}d\bm{k}_{2}d\bm{k}_{3}.
\label{eqn:exact_nk}
\end{equation}

The following derivation for the kinetic equation involves the standard procedure to relate the forth-order moment in \eqref{eqn:exact_nk} to second-order correlations through the quasi-Gaussian closure model. The detailed procedure can be found in  \cite{Majda1997,Zakharov2001,Nazarenko_WTTBOOK,Zakharov_WTTBOOK}. Taking a large time limit of the result (\eqref{eqn:exact_nk} with second-order correlations) yields the kinetic equation:

\begin{equation}
\begin{split}
\frac{\partial n_{\bm{k}}}{\partial t}=
\underset{\bm{k}_1,\bm{k}_2,\bm{k}_3 \in \mathbb{R}^2}{\int}4\pi k^{2} k_{1}^{2} k_{2}^{2} k_{3}^{2}
\left(n_{\bm{k}_{1}}n_{\bm{k}_{2}}n_{\bm{k}_{3}}+n_{\bm{k}_{1}}n_{\bm{k}_{2}}n_{\bm{k}}-n_{\bm{k}_{1}}n_{\bm{k}_{3}}n_{\bm{k}}-n_{\bm{k}_{2}}n_{\bm{k}_{3}}n_{\bm{k}}\right)
\\
\delta\left(\bm{k}_{1}+\bm{k}_{2}-\bm{k}_{3}-\bm{k}\right)\delta\left(\omega_{\bm{k}_{1}}+\omega_{\bm{k}_{2}}-\omega_{\bm{k}_{3}}-\omega_{\bm{k}}\right)d\bm{k}_{1} d\bm{k}_{2} d\bm{k}_{3}.
\label{eqn:KE_w_deltas}
\end{split}
\end{equation}
While a complete derivation of the Kolmogorov-Zakharov spectra (for both forward and inverse cascades) can be formulated \cite{Majda1997,Zakharov_WTTBOOK,Nazarenko_WTTBOOK}, only the spectral slope $\gamma_0$ for the  forward cascade is of interest for our present work. The value of $\gamma_0$ can be conveniently obtained through $\gamma_0=-2s/3-d$ (for quartet resonance) \cite{Nazarenko_WTTBOOK}, where $d$ is the dimension of $\bm{k}$ and $s$ is degree of homogeneity of the interaction kernel $V=kk_1k_2k_3$. With $d=2$ and $s=-\beta=4$ (for MMT), we obtain $\gamma_0=-14/3$.

\section{MMT formulations on a torus}
We define the Fourier series $\psi(\bm{x})=\sum_{\bm{k}\in\mathbb{Z}^2_{r,ir}} \hat{\psi}_{\bm{k}}e^{i\bm{k}\cdot\bm{x}}$ (for simplicity, we do not distinguish $\hat{\psi}_{\bm{k}}$ from the infinite domain notation), where $\mathbb{Z}^2_{r,ir} \equiv \mathbb{Z}\times q\mathbb{Z}$ with $q=1$ for $\mathbb{T}_r^2$ and $q^2=\sqrt{2}$ for $\mathbb{T}_{ir}^2$. The wave number domain formulation of \eqref{MMT} then yields
\begin{equation}
i\frac{\partial \hat{\psi}_{\bm{k}}}{\partial t}= k^{2}\hat{\psi}_{\bm{k}} +
\sum_{\substack{\bm{k}_{1},\bm{k}_{2},\bm{k}_{3} \in\mathbb{Z}^2_{r,ir}\\ \bm{k}_{1}+\bm{k}_{2}-\bm{k}_{3}-\bm{k}=0 }}{k k_{1} k_{2} k_{3}} \hat{\psi}_{\bm{k}_1}\hat{\psi}_{\bm{k}_2}\overline{\hat{\psi}_{\bm{k}_3}},
\label{eqn:MMT_dkdomain}
\end{equation}
Defining $S_{ext}\equiv (\{\bm{k}_{1},\bm{k}_{2},\bm{k}_{3})|\bm{k}_{1}+\bm{k}_{2}-\bm{k}_{3}-\bm{k}=0, \  \omega_{\bm{k}_{1}}+\omega_{\bm{k}_{2}}-\omega_{\bm{k}_{3}}-\omega_{\bm{k}}=0\}$, and $S_{qua}\equiv \{(\bm{k}_{1},\bm{k}_{2},\bm{k}_{3})|\bm{k}_{1}+\bm{k}_{2}-\bm{k}_{3}-\bm{k}=0, \  \omega_{\bm{k}_{1}}+\omega_{\bm{k}_{2}}-\omega_{\bm{k}_{3}}-\omega_{\bm{k}}\neq 0\}$, we split the summation in \eqref{eqn:MMT_dkdomain} into two terms
\begin{equation}
i\frac{\partial \hat{\psi}_{\bm{k}}}{\partial t}= k^{2}\hat{\psi}_{\bm{k}} +
\sum_{\substack{\bm{k}_{1},\bm{k}_{2},\bm{k}_{3} \in\mathbb{Z}^2_{r,ir} \\ (\bm{k}_{1},\bm{k}_{2},\bm{k}_{3})\in S_{ext}}}{k k_{1} k_{2} k_{3}} \hat{\psi}_{\bm{k}_1}\hat{\psi}_{\bm{k}_2}\overline{\hat{\psi}_{\bm{k}_3}}+
\sum_{\substack{\bm{k}_{1},\bm{k}_{2},\bm{k}_{3} \in\mathbb{Z}^2_{r,ir} \\ (\bm{k}_{1},\bm{k}_{2},\bm{k}_{3})\in S_{qua}}}{k k_{1} k_{2} k_{3}} \hat{\psi}_{\bm{k}_1}\hat{\psi}_{\bm{k}_2}\overline{\hat{\psi}_{\bm{k}_3}},
\label{eqn:MMT_dkdomain2}
\end{equation}

Considering $n_{\bm{k}}= \langle \hat{\psi}_{\bm{k}}\overline{\hat{\psi}_{\bm{k}}}  \rangle$ for the discrete case, the evolution of $n_{\bm{k}}$ can be derived (through similar procedures as the infinite domain case)

\begin{equation}
\frac{\partial n_{\bm{k}}}{\partial t}=
\sum_{\substack{\bm{k}_{1},\bm{k}_{2},\bm{k}_{3} \in\mathbb{Z}^2_{r,ir} \\ (\bm{k}_{1},\bm{k}_{2},\bm{k}_{3})\in S_{ext}}}
{2 k k_{1} k_{2} k_{3}}\text{Im}\langle\overline{\hat{\psi}_{\bm{k}}}\hat{\psi}_{\bm{k}_1}\hat{\psi}_{\bm{k}_2}\overline{\hat{\psi}_{\bm{k}_3}}\rangle +
\sum_{\substack{\bm{k}_{1},\bm{k}_{2},\bm{k}_{3} \in\mathbb{Z}^2_{r,ir} \\ (\bm{k}_{1},\bm{k}_{2},\bm{k}_{3})\in S_{qua}}}
{2 k k_{1} k_{2} k_{3}}\text{Im}\langle\overline{\hat{\psi}_{\bm{k}}}\hat{\psi}_{\bm{k}_1}\hat{\psi}_{\bm{k}_2}\overline{\hat{\psi}_{\bm{k}_3}}\rangle.
\label{eqn:dexact_nk}
\end{equation}

Understanding the summation over discrete $S_{ext}$ (DRM) in the first needs to leverage the \emph{continuous resonant} system, which will be introduced in the next section.

\section{Continuous Resonant System}

For the nonlinear Schrödinger equation $-i\partial_t a + \Delta a = \epsilon^2 |a|^2 a$ on $\mathbb{T}_L^2$ of size $L^2$, a theorem is proven in \cite{Faou2016} that at weakly nonlinear large box limits, a \emph{continuous resonant} equation holds
\begin{equation}
    -i\frac{\partial \tilde{a}_{\bm{k}}}{\partial t} = \frac{2\epsilon^2\log{L}}{\zeta(2)L^2} \int \tilde{a}_{\bm{k}_1}\tilde{a}_{\bm{k}_2}\overline{\tilde{a}_{\bm{k}_3}} \delta\left(\bm{k}_{1}+\bm{k}_{2}-\bm{k}_{3}-\bm{k}\right)\delta\left(\omega_{\bm{k}_{1}}+\omega_{\bm{k}_{2}}-\omega_{\bm{k}_{3}}-\omega_{\bm{k}}\right) d\bm{k}_{1} d\bm{k}_{2} d\bm{k}_{3},
    \label{eqn:Z-CR-EQ}
\end{equation}
where $\zeta$ is the Riemann zeta function, $\tilde{a}_{\bm{k}}=e^{-ik^2 t}\hat{a}_{\bm{k}}$, and $a(\bm{x})=L^{-2}\sum_{k\in\mathbb{Z}_L^2}\hat{a}_{\bm{k}}exp(2\pi i\bm{k}\cdot\bm{x})$ with $\mathbb{Z}_L^2 \equiv (L^{-1}\mathbb{Z})^2$ (note that this Fourier transform is different in normalization as that in part II, but consistent with \cite{Faou2016}).

Two important components in the derivation of \eqref{eqn:Z-CR-EQ} requires further clarification: (1) The weak nonlinearity limit is taken to eliminate the contribution of quasi-resonant interactions, so only resonant interactions are involved in \eqref{eqn:Z-CR-EQ}. This is observed in our numerical results and mathematically achieved by the normal form transformation; (2) The large box limit is taken such that the summation over $S_{ext}$ can be considered to be like a Riemann sum which is related to the continuous integration in \eqref{eqn:Z-CR-EQ}. For this purpose, the key equation established through number theory reads
\begin{equation}
    \sum_{\substack{\bm{k}_{1},\bm{k}_{2},\bm{k}_{3} \in\mathbb{Z}^2_{L} \\ (\bm{k}_{1},\bm{k}_{2},\bm{k}_{3})\in S_{ext}}}\hat{a}_{\bm{k}_1}\hat{a}_{\bm{k}_2}\overline{\hat{a}_{\bm{k}_3}} = \frac{2L^2\log{L}}{\zeta(2)} \underset{\substack{\bm{k}_1,\bm{k}_2,\bm{k}_3 \in \mathbb{R}^2 \\ (\bm{k}_1,\bm{k}_2,\bm{k}_3)\in S_{ext}}}{\int}\hat{a}_{\bm{k}_1}\hat{a}_{\bm{k}_2}\overline{\hat{a}_{\bm{k}_3}} d\bm{k}_{1} d\bm{k}_{2} d\bm{k}_{3}.
    \label{eqn:Z-CR}
\end{equation}

An alternative way to take the large box limit is to consider $k\rightarrow \infty$ for a given $L$ \cite{Faou2016}. Therefore, equation \eqref{eqn:Z-CR} can be reformulated for our configuration (changing from $\mathbb{Z}^2_{L}$ to $\mathbb{Z}^2_{r}$ and setting $\hat{a}_{\bm{k}}=k\hat{\psi}_{\bm{k}}$), leading to
\begin{equation}
    \sum_{\substack{\bm{k}_{1},\bm{k}_{2},\bm{k}_{3} \in\mathbb{Z}^2_{r} \\ (\bm{k}_{1},\bm{k}_{2},\bm{k}_{3})\in S_{ext}}}k_1k_2k_3\hat{\psi}_{\bm{k}_1}\hat{\psi}_{\bm{k}_2}\overline{\hat{\psi}_{\bm{k}_3}} \sim \underset{\substack{\bm{k}_1,\bm{k}_2,\bm{k}_3 \in \mathbb{R}^2 \\ (\bm{k}_1,\bm{k}_2,\bm{k}_3)\in S_{ext}}}{\int}k_1k_2k_3\hat{\psi}_{\bm{k}_1}\hat{\psi}_{\bm{k}_2}\overline{\hat{\psi}_{\bm{k}_3}} d\bm{k}_{1} d\bm{k}_{2} d\bm{k}_{3},
    \label{eqn:NUMPROB}
\end{equation}
where we have omitted the prefactor (the treatment of which will involve more subtlety and is not necessary for the purpose of this paper). Equation \eqref{eqn:NUMPROB} can then be used to replace the summation over $S_{ext}$ on $\mathbb{T}_r^2$ in \eqref{eqn:MMT_dkdomain2} into an integral. Under low nonlinearity, we arrive at
\begin{equation}
\frac{\partial n_{\bm{k}}}{\partial t} = \frac{\partial n_{\bm{k}}}{\partial t} \bigg|_{ext} \sim 
\underset{\substack{\bm{k}_1,\bm{k}_2,\bm{k}_3 \in \mathbb{R}^2 \\ (\bm{k}_1,\bm{k}_2,\bm{k}_3)\in S_{ext}}}{\int} 2 k k_{1} k_{2} k_{3} \text{Im}\langle\overline{\hat{\psi}_{\bm{k}}}\hat{\psi}_{\bm{k}_1}\hat{\psi}_{\bm{k}_2}\overline{\hat{\psi}_{\bm{k}_3}}\rangle d\bm{k}_{1}d\bm{k}_{2}d\bm{k}_{3}, \text{on}\ \mathbb{T}_r^2.
\label{eqn:CR_supp}
\end{equation}

\section{Resonant Quartets on $\mathbb{T}^2_r$ and $\mathbb{T}^2_{ir}$}

We consider the exact resonance condition for a quartet:
\begin{equation}
\begin{split}
\bm{k}_{1}+\bm{k}_{2}-\bm{k}_{3}-\bm{k}&=0, \\
\omega_{\bm{k}_{1}}+\omega_{\bm{k}_{2}}-\omega_{\bm{k}_{3}}-\omega_{\bm{k}}&=0,
\label{eqn:res_cond}
\end{split}
\end{equation}
with dispersion relation $\omega_{\bm{k}}=\vert \bm{k} \vert^{2}$. For discrete wave number $\bm{k}=(n,q m)$ with $m,n \in \mathbb{Z}$ on $\mathbb{T}_r^2$ ($q=1$) and $\mathbb{T}_{ir}^2$ ($q^2=\sqrt{2}$), the resonance condition \eqref{eqn:res_cond} can be expanded as
\begin{align}
n_{1}+n_{2}-n_{3}-n&=0
\label{eqn:disc_res_cond1}
\\
m_{1}+m_{2}-m_{3}-m&=0
\label{eqn:disc_res_cond2}
\\
n_{1}^{2}+n_{2}^{2}-n_{3}^{2}-n^{2}&=-q^{2}\left(m_{1}^{2}+m_{2}^{2}-m_{3}^{2}-m^{2}
\right)
\label{eqn:disc_res_cond3}
\end{align}
Substituting (\ref{eqn:disc_res_cond1}) and (\ref{eqn:disc_res_cond2}) into (\ref{eqn:disc_res_cond3}) yields
\begin{equation}
(n_{1}-n)(n_{2}-n)=-q^{2}(m_{1}-m)(m_{2}-m)
\label{eqn:disc_cond_comb}
\end{equation}

\begin{figure}[h!]
\includegraphics[width=8.6cm]{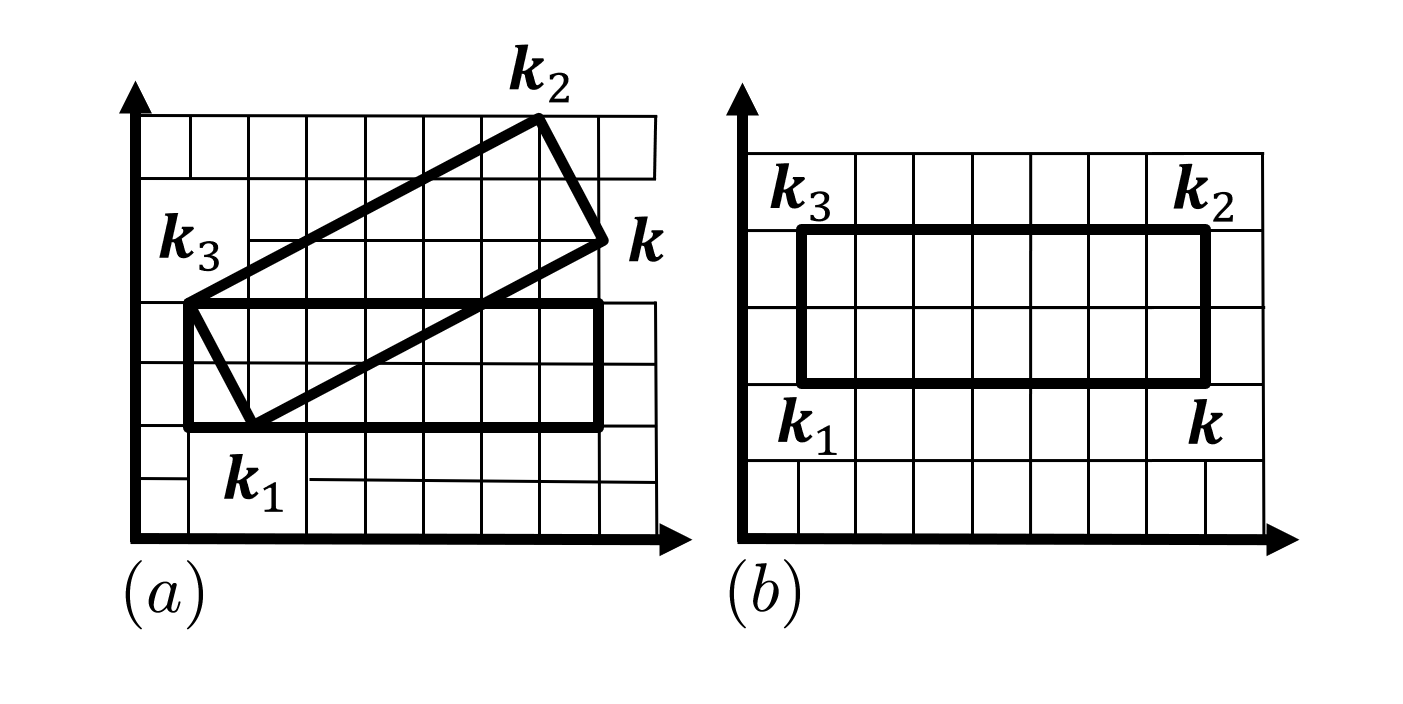}
\caption{Sketches of the quadrilaterals formed by vertices $\bm{k}$, $\bm{k}_1$, $\bm{k}_2$ and $\bm{k}_3$ on (a) $\mathbb{T}_r^2$ and (b) $\mathbb{T}_{ir}^2$.}
\label{fig:rect}
\end{figure}

For $\mathbb{T}^2_r$ ($q=1$), \eqref{eqn:disc_cond_comb} is reduced to $(\bm{k}_{1}-\bm{k})\cdot(\bm{k}_{2}-\bm{k})=0$, i.e., $(\bm{k}_{1}-\bm{k})\perp(\bm{k}_{2}-\bm{k})$. Combined with \eqref{eqn:disc_res_cond1} and \eqref{eqn:disc_res_cond2}, it can be understood that the four vertices represented by $\bm{k}$, $\bm{k}_1$, $\bm{k}_2$ and $\bm{k}_3$ form a quadrilateral with arbitrary orientations allowed on the discrete wave number grid (figure \ref{fig:rect}(a)). For $\mathbb{T}^2_{ir}$ ($q^2=\sqrt{2}$), \eqref{eqn:disc_cond_comb} holds only if LHS=RHS=0, i.e., $n\in\{n_{1},n_{2}\}$ and $m\in\{m_{1},m_{2}\}$. Therefore, the four vertices  $\bm{k}$, $\bm{k}_1$, $\bm{k}_2$ and $\bm{k}_3$ form a quadrilateral with only horizontal/vertical orientations allowed, i.e., aligned with axes (figure \ref{fig:rect}(b)). 

\begin{figure}[h!]
\centering
\subfigure{\includegraphics[width=6.3cm]{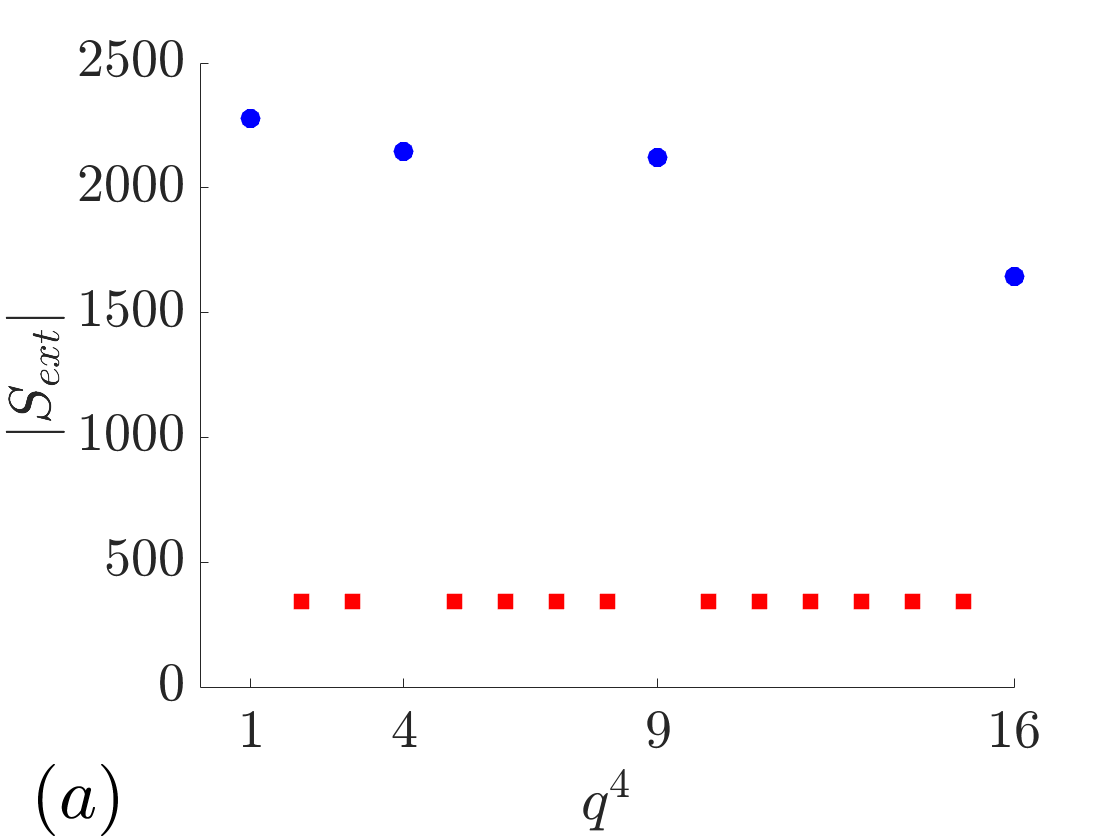}}
\subfigure{\includegraphics[width=6.3cm]{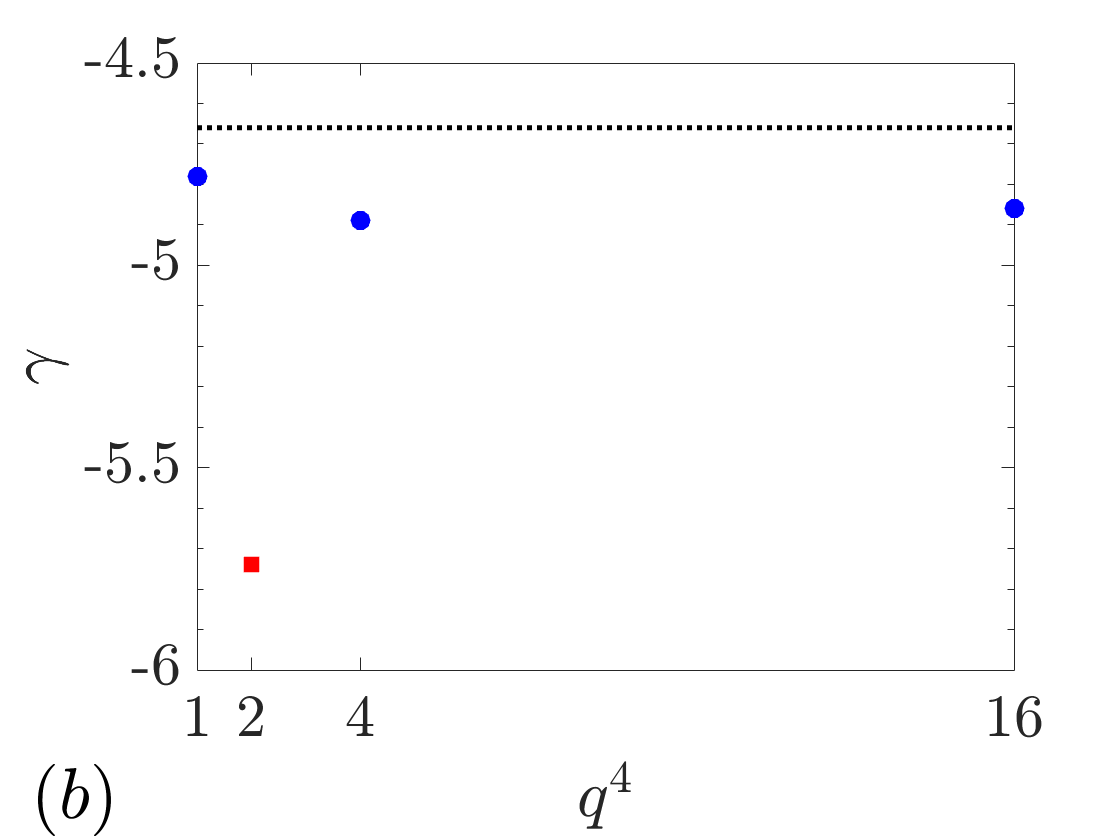}}
\caption{(a) $|S_{ext}|$ computed for rational ($\bm{\bullet}$) and irrational (\tiny\mbox{$\blacksquare$}\normalsize) $q^2$, with $\bm{k}_2=(-36,31q)$ and $k_{3x}=-22$ as considered in the main paper. (b) Spectral slope $\gamma$ computed using $\nu_{opt}$ with $N=1024^2$ modes for rational ($\bm{\bullet}$) and irrational (\tiny\mbox{$\blacksquare$}\normalsize) $q^2$. The WTT analytical solution $\gamma_0=-14/3$ is indicated (\dotL).}
\label{fig:card_s}
\end{figure}

We further provide a quantitative study on the effect of $q$ to the number of exact resonances (i.e., the sparsity/density of resonances on the DRM). For $\bm{k}$, $\bm{k}_1$, $\bm{k}_2$, $\bm{k}_3$ $\in([-M,M],q[-M,M])$ with $M=85$, we plot in figure \ref{fig:card_s}(a) the cardinality $|S_{ext}|$, i.e., number of elements in $S_{ext}$, for $q^4\in \{n|n\in \mathbb{Z}, 1\leq n \leq 16 \}$. We see that when $q^2$ is irrational (red points), $|S_{ext}|$ is minimized due to the reason associated with figure \ref{fig:rect}. For rational $q^2$, $|S_{ext}|$ is significantly larger with the maximum attained at $q=1$ (in the test range). This can be understood in terms of the number of solutions of \eqref{eqn:disc_cond_comb} which is reduced due to the constrain in the prime factors introduced by $q\neq 1$. Therefore, our choices of $q^2=\sqrt{2}$ and $q=1$ are the two extremes in terms of $|S_{ext}|$. 

The spectral slopes for four selected values of $q$ are shown in figure \ref{fig:card_s}(b). These include the existing two cases of $q^2=\sqrt{2}$ and $q=1$, with additional two cases of $q=\sqrt{2}$ and $q=2$. It can be seen that the case with $q^2=\sqrt{2}$ leads to the largest deviation of $\gamma$ to the WTT solution $\gamma_0$. Among the other three values of $q$, $\gamma(q=1)$ is closest to $\gamma_0$, consistent with the largest number of exact resonances. However, $\gamma(q=2)$ is slightly closer to $\gamma_0$ than $\gamma(q=\sqrt{2})$ even though considerably more resonances are identified for the case with $q=\sqrt{2}$. This suggests that the ``sparsity'' of the DRM is not the only factor to consider for the DWT dynamics, but that the distribution of the resonances also plays a role. A criterion to measure the ``closeness'' of the DWT dynamics to WTT may be established through the \emph{continuous resonant} equation, which is only proved for the $q=1$ case. The discrete wave numbers of $q=1$ align periodically with those in $q=2$, but never overlaps with those in $q=\sqrt{2}$. This may lead to the spectral slope $\gamma$ for $q=2$ being closer to $\gamma_0$ even though fewer exact resonances are present than in the case of $q=\sqrt{2}$. The argument may be generalized to the comparison between (a) rational $q$ and (b) irrational $q$ but rational $q^2$ in the large wave number limit, but more theoretical and numerical studies are required.





\vspace{-0.8cm}
\section{$k-\omega$ spectral analysis and quasi-DRM visualization}
\vspace{-0.4cm}

\begin{figure}
  \centering
  \subfigure{\includegraphics[width=6.0cm]{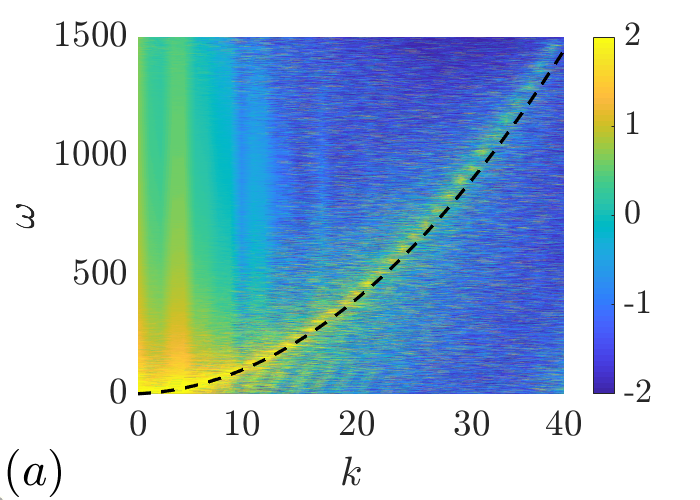}}
  \subfigure{\includegraphics[width=6.0cm]{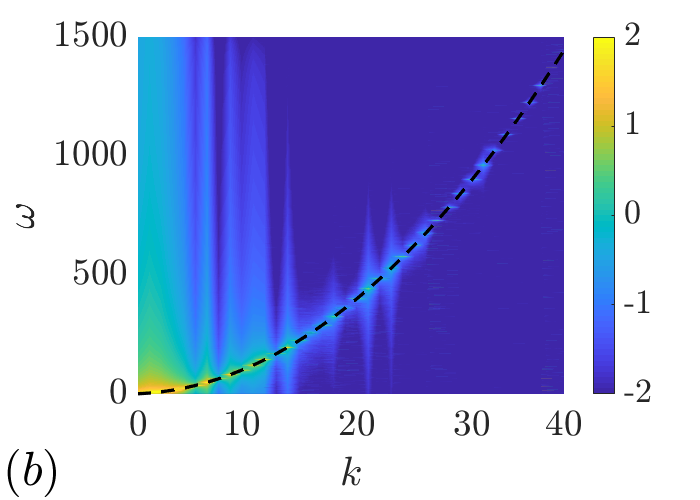}}
  \subfigure{\includegraphics[width=6.0cm]{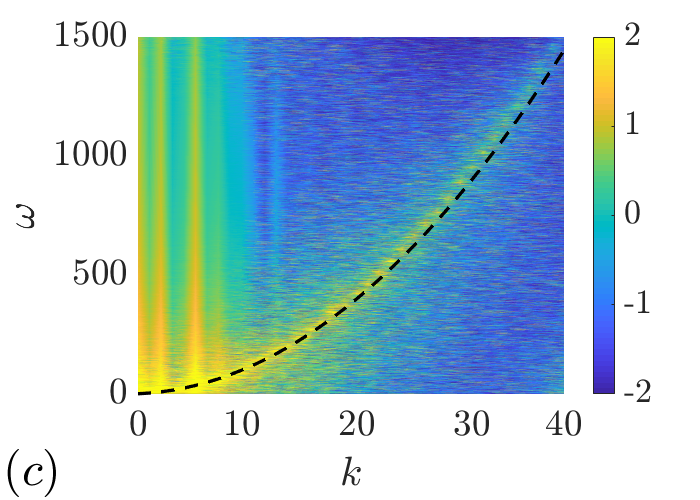}}
  \subfigure{\includegraphics[width=6.0cm]{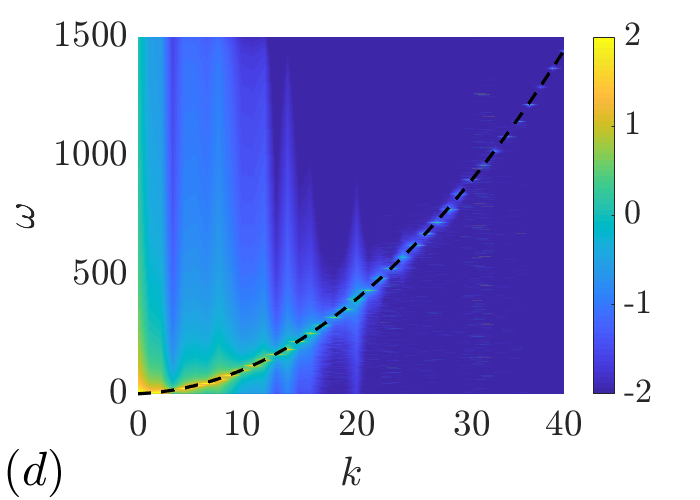}}
  \caption{$\omega-k$ spectra on $\mathbb{T}^2_{r}$ for (a) $\epsilon = 3\times10^{-2}$ and (b) $\epsilon = 3\times10^{-4}$, and on $\mathbb{T}^2_{ir}$ for (c) $\epsilon = 3\times10^{-2}$ and (d) $\epsilon = 3\times10^{-4}$. The dispersion relation $\omega=k^2$ is indicated $(\dashL)$.}
  \label{fig:k-w}
\end{figure}

We plot the (angle-averaged) $k-\omega$ spectra $E(k,\omega)$ on $\mathbb{T}^2_{r}$ and $\mathbb{T}^2_{ir}$ for $\epsilon=3.00\times10^{-2}$ and $\epsilon=3.00\times10^{-4}$ in figure \ref{fig:k-w}. We confirm that most high energy density is concentrated near the linear dispersion relation $\omega=k^2$ for all the cases. While some bound wave components exist at higher nonlinearity level, they are insignificant for lower nonlinearity level (and similar on $\mathbb{T}^2_{r}$ and $\mathbb{T}^2_{ir}$). Therefore, bound waves cannot be used to interpret the spectral slope difference between $\mathbb{T}^2_{r}$ and $\mathbb{T}^2_{ir}$. The nonlinear frequency broadening $\Omega$ can be estimated for all cases by taking the width of $E(k,\omega)$ around $\omega=k^2$ at $k=25$ in the inertial range (without loss of generality). Measuring the width by covering up to the $50\%$ of the maximum value \cite{Boue2011}, we obtain the nonlinear frequency broadening $\Omega^+\approx10$ and $\Omega^-\approx1$ for higher and lower nonlinearity levels on both $\mathbb{T}^2_{r}$ and $\mathbb{T}^2_{ir}$.  

\begin{figure}[h!]
  \centering
  \subfigure{\includegraphics[width=5.9cm]{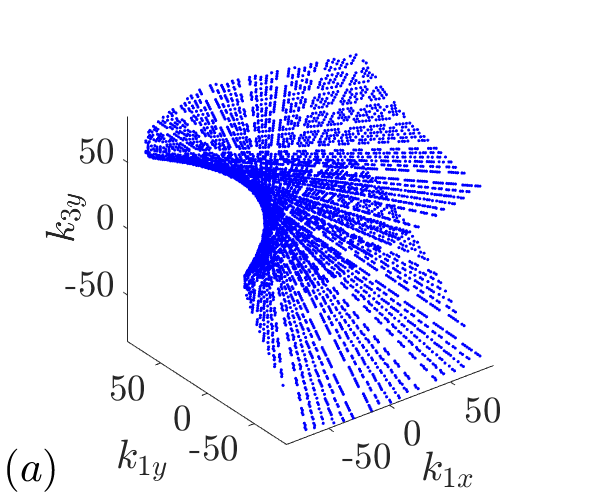}}\quad
  \subfigure{\includegraphics[width=5.9cm]{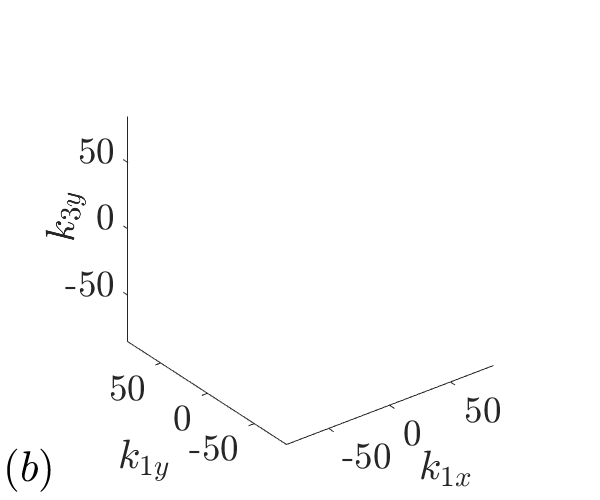}}
  \subfigure{\includegraphics[width=5.9cm]{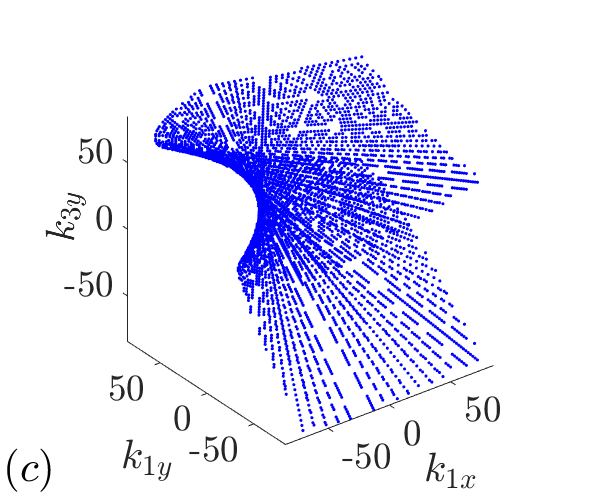}}\quad
  \subfigure{\includegraphics[width=5.9cm]{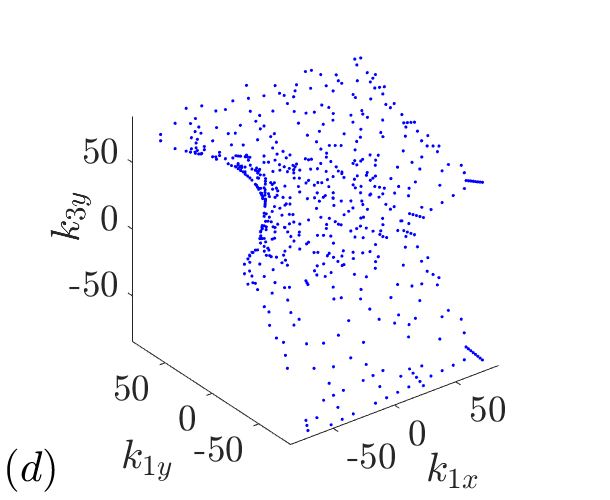}}
  \caption{$S_{\Omega^\pm}\setminus S_{ext}$ on $\mathbb{T}^2_{r}$ for (a) $\Omega^+=10$ and (b) $\Omega^-=1$, and on $\mathbb{T}^2_{ir}$ for (c) $\Omega^+=10$ and (d) $\Omega^-=1$.}
  \label{fig:qdrm}
\end{figure}

We define the quasi-DRM($\Omega^\pm$) as the set $S_{\Omega^\pm}\setminus S_{ext}$, where $S_{\Omega^\pm}\equiv \{(\bm{k}_{1},\bm{k}_{2},\bm{k}_{3})|\bm{k}_{1}+\bm{k}_{2}-\bm{k}_{3}-\bm{k}=0, \  |\omega_{\bm{k}_{1}}+\omega_{\bm{k}_{2}}-\omega_{\bm{k}_{3}}-\omega_{\bm{k}}|<\Omega^\pm\}$. By setting $\bm{k}_2=(-36,31q)$ and $k_{3x}=-22$, figure \ref{fig:qdrm} shows the quasi-DRM($\Omega^\pm$) on both $\mathbb{T}^2_{r}$ and $\mathbb{T}^2_{ir}$. On $\mathbb{T}^2_{r}$, we find an empty set for lower nonlinearity level, indicating the dominance of exact resonance, in contrast to the quasi-DRM at higher nonlinearity level which provides a considerable portion of $P$. On $\mathbb{T}^2_{ir}$, the quasi-DRM at higher nonlinearity is significantly ``denser'' than that at lower nonlinearity, resulting in the variation of spectral slope and energy cascade with $\epsilon$ observed in the paper.


\newpage
\bibliography{citations}